\documentclass[12pt]{article} 
\usepackage[dvips]{graphics}
\begin{document} 
		
\title{On the Very-High-Energy Gamma Ray Spectra from Typical Supernovae Remnants}  
\author{{\it Richard Shurtleff~}\thanks{affiliation and mailing 
address: Department of Applied Mathematics and Sciences, 
Wentworth Institute of Technology, 550 Huntington Avenue, 
Boston, MA, USA, ZIP 02115, telephone number: (617) 989-4338, e-mail address: shurtleffr@wit.edu}} 
          
\maketitle 

\begin{abstract} Recently measured VHE gamma ray spectra from supernovae remnants (SNRs) are best fit by power laws with an exponential cutoff. But this feature does not occur at the `knee-equivalent' energy at which VHE gamma ray spectra are expected to reflect the `knee' in the otherwise featureless cosmic ray spectrum. In this article, the VHE gamma ray feature is explained as a consequence of a recently deduced quantum effect. The effect distinguishes `trajectory energy' from `particle state energy' and requires that the particle state energy depends strongly on gravitational potential at very high energies. Based on this effect and the observed CR spectrum, a tight two-parameter fit is obtained to the combined VHE gamma ray spectra of SNRs RX J1713.7-3946 and RX J0852.0-4622 and the Galactic Center ridge. 


\end{abstract}

\pagebreak
\section{Introduction} \label{intro}

It is widely held that galactic cosmic rays are mostly atomic nuclei accelerated in supernovae remnants, that some CRs collide with local ambient matter producing gamma rays, and that CRs, being charged, are deflected as they travel to Earth. Being uncharged, the gamma rays that arrive at Earth retain directional information pointing back to their sources.

\begin{figure}[ht]  
\centering
\vspace{0cm}
\hspace{0in}\includegraphics[0,0][360,360]{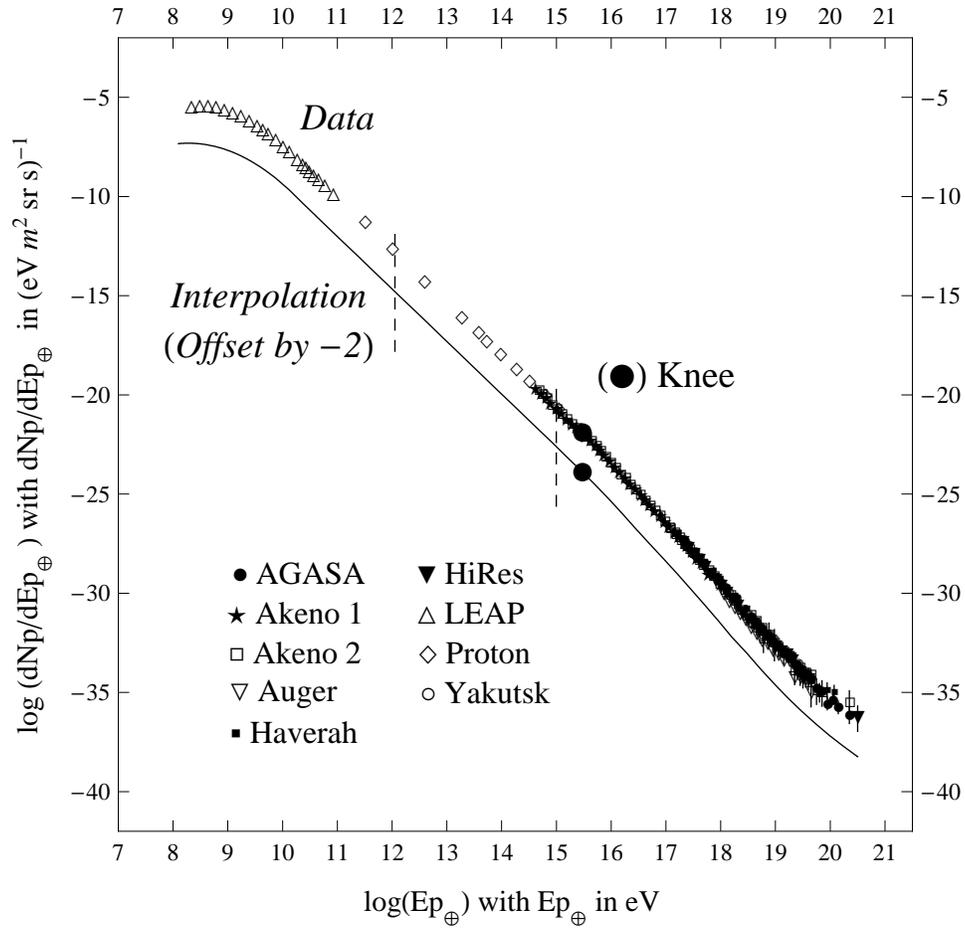}
\caption{{\it{The Cosmic Ray Spectrum at Earth.}} The dashed lines indicate the region of the observed cosmic ray (CR) spectrum that corresponds to the observed VHE gamma ray spectrum. In that region the CR spectrum is a near-pure power law without any distinctive feature. The `Knee' in the CR spectrum, where the spectral index changes, occurs at a higher energy and should have little effect on the observed VHE gamma ray spectrum.  }
\label{CRearth}
\end{figure}

Since the galactic CR primaries have been deflected, the CR spectrum observed at Earth, see Fig. \ref{CRearth}, represents an amalgam of SNR sources. The spectrum follows a near-constant power law from $10^{10}$ eV to the `knee' at $3\times 10^{15}$ eV, where the spectral index of the power law steepens and is again constant to the `ankle' at $10^{18}$ eV. The `ultrahigh energy' CRs do not concern us here. The most energetic VHE gamma rays thus far observed have energies $10^{14}$ eV or so, implying that the relevant CR energies are below about $10^{15}$ eV. 

Considering only proton primaries, CR protons at the SNR are expected to produce VHE gamma rays when the protons collide with ambient matter, most likely protons. The $p$-$p$ collisions create pions and etas that decay to produce the VHE gamma rays. The process is well understood and the only quantity that could provide the VHE gamma ray spectra with a distinct feature is the inelastic $p$-$p$ cross section $\sigma_{inel}.$ 

However the log of $\sigma_{inel}$ is a slowly rising function of incident proton energy and is remarkably featureless, see Fig. \ref{sigmadEdE}a. Thus one expects the VHE gamma ray spectra from CR protons to be a power law up to some `knee equivalent' gamma ray energy, and at higher energies to again be a power law but with a probably steeper index. In short, the VHE gamma ray spectrum should be similar to the CR spectrum and a typical VHE spectrum should feature a knee equivalent somewhere around $0.17\times 3 \times 10^{15}$ =  $5.1\times 10^{14}$ eV. 

The H.E.S.S. collaboration reports a cutoff in the spectrum for the supernova remnant RX J1713.7-3946 (Aharonian et al. 2007a) at an energy of at most $1.8 \times 10^{13}$ eV, a factor of about 30 below the expected knee equivalent. Therefore, it is thought that the feature reported is not the knee equivalent (Plaga 2008a). One must consider other explanations. It could be that the VHE gamma rays are emitted by electrons in the `lepton scenario' since electrons are also believed to be accelerated in supernovae remnants and electrons in certain conditions emit VHE gamma rays. For lepton scenario discussions, see Huang et al (2007), Katz \& Waxman (2008), and Plaga (2008b). 

It is shown in this paper that the hadron scenario fits the VHE gamma ray feature at $10^{13}$ eV if we apply the result of a recent calculation. The explanation distinguishes two different kinds of energy that are equal at low energies and diverge at very high energies. The `particle state energy' depends on gravitational potential, with less energy at the SNR and more energy at Earth (Shurtleff 2008). A summary of the calculation is included in Sec.~\ref{PDPSE} for convenience.

Then a massive particle such as a proton has two energies, one that can be obtained by mass and a time-of-flight velocity, a `trajectory energy'. The other is the energy delivered in a collision due to the change in the particle's state. Call that the `particle state energy'.   The formula for the particle state energy $E$ is
\begin{equation} \label{Eintro}
E = E_{0} -4 \phi \frac{E_{0}^{3}}{m^2 c^4} \, ,
\end{equation}
where $\phi < 0$ is the gravitational potential and  $E_{0}$ is the energy at null potential $\phi$ = 0. The gravitational field is assumed to be weak and the potential $\phi$ is the Newtonian potential made unitless by dividing by the square of the speed of light $c.$ The mass of the particle is $m.$

The Solar System moves in the gravitational field of the Galaxy. Very roughly, the Galaxy should contribute a potential $\phi_{Gal}$ due to $M_{Gal}$ = $10^{11}$ solar masses in effect collected some $R$ = $2.5\times10^{20}$ m away from the Solar System at the Galactic Center. One has $\phi_{Gal}$ = $-GM_{Gal}/Rc^{2} \approx$ $-6\times 10^{-7}.$  It follows that the effects of Eq. (\ref{Eintro}) would be significant at about $E_{0} $ = $mc^{2}/2\sqrt{-\phi_{Gal}}$ = 0.6 TeV, where $mc^{2} \approx$ 0.001 TeV is the rest energy of a proton. Since 0.6 TeV is well within the range of proton energies available at the Tevatron, and no such effects are reported from the Tevatron, the highest energy proton accelerator as of this writing, it follows that the potential due to the Galaxy and any larger potentials must be eliminated from consideration if Eq. (\ref{Eintro}) is true for protons.
 
To eliminate the effects of the Galaxy's gravitational field, one may choose to work in a comoving coordinate system in free fall with the Solar System, see for example Weinberg (1972). In General Relativity, a particle's trajectory can be described in any coordinate system. 

By making the particle state energy dependent on potential and requiring the potential to be calculated in a system in free fall with the Solar System, the description is valid in one coordinate system, the reference frame of the Solar System. This is allowed in General Relativity as long as the various quantities are properly transformed when the description is given in some other coordinate system.  

While the coordinate system is in free fall with the Solar System, thereby canceling the Galactic and larger scale potentials, the local gravitational potential due to the matter in the Solar System does not cancel. Thus, at the Earth's surface, the $\phi$ in (\ref{Eintro}) is the potential due to the Sun and the Earth, a value of about $\phi_{\oplus} \approx$ $-1\times 10^{-8},$ which becomes significant at an energy $E_{0}\approx$ 5 TeV. This is beyond the proton energies available at proton accelerators at the time of this writing, but is within the reach of the LHC which has not begun high energy runs at the time of this writing.

While the properties of electrons have no bearing on the explanation of the feature in the VHE gamma ray spectrum, one may note the energy at which the effects of (\ref{Eintro}) would become significant for electrons is $E_{0} $ = $m_{e}c^{2}/2\sqrt{-\phi_{\oplus}}$ = 2.5 GeV. Since the effects expected by applying (\ref{Eintro}) have not been reported by accelerator experiments with electron energies many times the threshold 2.5 GeV, it follows that electrons do not obey (\ref{Eintro}), at least for potentials $\phi$ with magnitudes as large as the potential due to the Earth and Sun. This could mean the comoving coordinate system for electrons is in free fall with matter falling in the gravitational field due to the Earth and Sun (and due to the Galaxy and larger scales as well). Or it could mean that the assumptions that produce (\ref{Eintro}) do not apply to electrons. This could be another way that electrons differ from protons.

One benefit of going to a system in free fall with the local matter is the resulting universality possible for the various SNRs. If the potential at each SNR includes the contribution from the Galaxy, then Eq. (\ref{Eintro}) would have a different $\phi_{S}$ for each supernova remnant. That would be a troubling problem. But if the potential $\phi_{S}$ is the local potential in the system in free fall with the SNR, then the potentials at the various SNRs are likely to be nearly the same and Eq. (\ref{Eintro}) can meaningfully be applied to SNRs anywhere in the Galaxy.  

In this picture, to get a cosmic ray from the SNR to Earth, one must consider a number of overlapping patches of spacetime, each patch described by a coordinate system in free fall with the local matter at a scale comparable to objects like the Solar System or supernova remnants.

With Eq. (\ref{Eintro}), a cosmic ray proton observed at Earth with energy $Ep_{\oplus}$ in the potential $\phi_{\oplus}$ at the Earth has a different energy $Ep_{S}$ at the supernova in the potential $\phi_{S}.$ In this notation, `$p$' stands for proton and `$\oplus$' for Earth and `$S$' for supernova remnant. 

Relating the CR spectrum at Earth,  $dN_{p}/dEp_{\oplus},$ to the CR spectrum at a typical SNR, $dN_{p}/dEp_{S},$ involves the derivative $dEp_{\oplus}/dEp_{S}$ relating the energy at the SNR and the corresponding energy at Earth. It is suggested here that this energy dependence is part of the feature found by the H.E.S.S. collaboration in the VHE spectrum of SNR RX J1713.7-3946. 

Having obtained the expected CR spectrum at a typical SNR based on the CR spectrum at Earth, the expected VHE gamma ray spectrum at the SNR can be determined by the analysis by Kelner et al. (2006). Then the expected VHE gamma ray spectrum at Earth can be found if we know how it is related to the VHE gamma ray spectrum at the SNR.

The problem is how to relate the VHE gamma ray spectrum at the SNR to the VHE gamma ray spectrum at Earth. Since the basic equation (\ref{Eintro}) applies to massive particles, one may entertain various arguments to get it to apply to massless gamma rays. 

First of all, note that at 10 TeV, say, a proton's rest energy is just 0.01\% of its total energy. The proton moves very much like a massless photon of the same energy. So a 10 TeV VHE gamma ray should behave much like a 10 TeV proton. Since the basis of the effect separates this `trajectory energy' from the `particle state energy', this argument may be spurious, even though it is intuitively appealing.

Secondly, consider the process of neutral pion decay that produces many of the VHE gamma rays in the hadronic scenario. A 10 TeV neutral pion decays to a 10 TeV VHE gamma ray plus a 4 keV X-ray. Within roundoff, the pion and the gamma ray each have an energy of 10 TeV. By time inversion, starting with a 10 TeV VHE gamma ray, an interaction with a 4 keV X-ray could produce a 10 TeV neutral pion. However unlikely the process, it could happen, so a 10 TeV VHE gamma ray is virtually a 10 TeV pion. Since the virtual transition from VHE gamma ray to neutral pion could occur anywhere from the SNR to Earth and a neutral pion traveling from the SNR to Earth would obey Eq. (\ref{Eintro}), the energy of the VHE gamma should obey Eq. (\ref{Eintro}) with $m$ as the neutral pion mass $m_{\pi^{0}}.$ 

Imagine placing a shuttered X-ray source at each kilometer along the path of a VHE gamma ray from an SNR to Earth. Furthermore let there be a limitless supply of identical VHE gamma rays ready to be sent from the SNR to Earth one by one. Let the $n$ = 0 VHE gamma ray travel from the SNR to Earth without removing any of the shutters. This VHE gamma ray reaches Earth unchanged by the presence of the X-ray sources. For the next VHE gamma ray, i.e. $n$ = 1, release X-rays at the first kilometer at the SNR so that a neutral pion is created by inverse pion decay. The neutral pion has almost the same energy as the VHE gamma ray. For the $n$th VHE gamma ray release X-rays from just the one source at the $n$th kilometer making an $n$th neutral pion which has almost the same energy as the $n$th VHE gamma ray at the $n$th kilometer. Since the $n$ = 0 VHE gamma ray has the same energy as the $n$th VHE gamma ray just before the $n$th kilometer, and the $n$th VHE gamma ray has nearly the same energy as a neutral pion at the $n$th kilometer, and a neutral pion obeys (\ref{Eintro}), it follows that the VHE gamma ray energy should also obey Eq. (\ref{Eintro}) with $m$ equal to the mass of a neutral pion. 

It should be clear that the result does not depend on there being any real X-rays along the path of the VHE gamma ray from the SNR to Earth. It should also be clear that none of this implies that a VHE gamma ray has mass, only that its energy could be transferred essentially undiminished to a neutral pion in some possibly imaginary circumstances.

But there are other ways a 10 TeV VHE gamma ray could interact with ambient matter of negligible energy, perhaps making the VHE gamma ray virtually an electron or a positron or an eta meson, etc. Again, there need not be any real ambient matter, imaginary ambient matter suffices. Not knowing how to calculate the likelihood of independent imaginary interactions means not knowing what mass to associate with the VHE gamma ray. Thus we let that mass be a free parameter. 

By these arguments, we apply Eq. (\ref{Eintro}) to VHE gamma rays by letting the mass $m$ = $m_{ray}$ stand for the average mass of the masses of massive particles that could be given the same energy as the VHE gamma ray by some possibly imaginary process. The mass $m_{ray}$ is a free parameter found by fitting the experimental VHE gamma ray spectra. 

Allowing (\ref{Eintro}) to apply to the VHE gamma rays allows us to use the expected VHE gamma ray flux at the SNR, $dN_{\gamma}/dE\gamma_{S},$  to determine the expected VHE gamma ray flux at Earth, $dN_{\gamma}/dE\gamma_{\oplus}.$ The relationship involves the derivative $dE\gamma_{S}/dE\gamma_{\oplus}$ which is energy dependent. 

As discussed previously, one also has the derivative $dEp_{\oplus}/dEp_{S}$ from the CR spectrum. The product of the two factors, $dEp_{\oplus}/dEp_{S}$ times $dE\gamma_{S}/dE\gamma_{\oplus},$ is an energy dependent quantity, see Fig. \ref{sigmadEdE}b, that could describe the feature in the VHE gamma ray spectrum of the SNR RX J1713.7-3946 reported by the H.E.S.S. collaboration.

\section{Potential-Dependent Particle State Energy} \label{PDPSE}

This section is included to indicate the origin of the basic formula. For details see Shurtleff (2008).

The quantum field for a massive particle species can be expanded over the annihilation and creation operators that remove or add particle states to  multiparticle states, see, for example, Weinberg (1995). Consider an annihilation field $\psi^{+},$  
\begin{equation} \label{psi+}
\psi^{+}_{l}(x) = \sum_{\sigma} \int d^3 p \enspace u_{l\sigma}(x,{\overrightarrow{p}}) a_{\sigma}({\overrightarrow{p}})  \, ,
\end{equation}
where the annihilation operator $a_{\sigma}({\overrightarrow{p}})$ removes a state of the particle with (four-)momentum $\{ \overrightarrow{p},p^{t} \}$ with total energy $p^{t}$ = $+\sqrt{m^{2} - \overrightarrow{p}^{2}}.$ The $u_{l\sigma}(x,{\overrightarrow{p}})$ are the `coefficient functions.' 

To comply with special relativity, the quantum field $\psi^{+}_{l}(x)$ transforms linearly under a Poincar\'{e} transformation $\{\Lambda,b\},$ where $\Lambda$ is the result of successive rotations and boosts, a `Lorentz transformation', followed by a translation along the displacement $b.$ Such transformations preserve the scalar products of coordinate differences between events. Likewise the operator $a_{\sigma}({\overrightarrow{p}})$ transforms linearly with a Poincar\'{e} transformation.

An essential step in obtaining the basic formula needed in this article depends on a characteristic of Poincar\'{e} transformations, a property of translations that is not shared with rotations and boosts. Unlike with rotations and boosts, under a translation the coordinate differences themselves are invariant. Thus it is impossible to tell by keeping track of coordinate differences just what the displacement was. One displacement $\epsilon$ could have been applied to the annihilation operator and a displacement $b$ could have been applied to the quantum field. The displacements $\epsilon$ and $b$ may be independent of one another without any effect on the underlying coordinate differences. 

The annihilation and creation operators and the particle states must transform via unitary reps to satisfy fundamental quantum principles. The transformation $\{\Lambda,\epsilon\}$ applied to the annihilation operators can be written as follows,
\begin{equation} \label{Da+}
U(\Lambda,\epsilon) a_{\sigma}({\overrightarrow{p}}) {U}^{-1}(\Lambda,\epsilon) = e^{i \Lambda p \cdot \epsilon} \sqrt{\frac{(\Lambda p)^t}{p^t}}  \sum_{\bar{\sigma}} {D^{(j)}_{\sigma \bar{\sigma}}}^{-1}  a_{\bar{\sigma}}({\overrightarrow{\Lambda p}}) \, ,
\end{equation}
where $D^{(j)}$ is the transformation matrix and $j$ is the spin of the particle. What is important here is the phase factor, $e^{i \Lambda p \cdot \epsilon},$ which is a unitary representation of $\{\Lambda,\epsilon\}.$ 

With the transformation $\{\Lambda,\epsilon\}$ applied to the annihilation operators, the quantum field is required to transform as follows, 
\begin{equation} \label{Dpsi}
U(\Lambda,b) \psi^{+}_{l}(x) {U}^{-1}(\Lambda,b) = \sum_{\bar{l}} D^{-1}_{l \bar{l}}(\Lambda,b)  \psi^{\pm}_{\bar{l}}(\Lambda x + b) \, ,
\end{equation}
where $\Lambda x + b$ results from transforming the coordinates and $D(\Lambda,b)$ is the nonunitary transformation matrix corresponding to the possibly reducible spin, say  $(A,B)\oplus(C,D)\oplus \ldots ,$ of the field $\psi^{+}.$ For a spin $j$ = 1/2 Dirac particle the field transforms with spin $(1/2,0)\oplus(0,1/2).$ 

Thus one object, the quantum field $\psi^{+},$ is a linear combination of other objects, the operators $a_{{\sigma}}({\overrightarrow{ p}}),$ that transform differently from the first type of object. In such cases, the coefficients, here the $u_{l\sigma}(x,{\overrightarrow{p}}),$ are constrained. Clebsch-Gordon coefficients can be derived from such constraints involving rotations in Euclidean 3-space (Shankar 1994). Here the transformations differ in part because one is unitary and the other nonunitary. Also the displacement for the operators $\epsilon$ may differ from the displacement $b$ for the quantum field. Deriving free quantum fields from special relativity in this way is presented in Chapter 5 of Weinberg (1995).

In order to obtain the basic formula, consider the arbitrary displacement $\epsilon.$ Since $\epsilon$ is arbitrary, it may be useful to restrict it somewhat and make it depend on various relevant quantities. Let $\epsilon$ depend on the field's Poincar\'{e} transformation $\{\Lambda,b \}$ and also depend on the event coordinates $x.$ The constraints that determine the coefficient functions $u_{l\sigma}(x,{\overrightarrow{p}})$ also constrain the function $\epsilon(\Lambda,x,b).$ 

Leaving the details to the references, one finds that the coefficient functions are determined by (\ref{psi+}), (\ref{Da+}), (\ref{Dpsi}) to be
\begin{equation} \label{u}
  u_{l\sigma}(x ,{\overrightarrow{p}}) = \sqrt{\frac{ m}{p^{\, t}}} \, e^{i p \cdot Mx} \sum_{\bar{l}} D_{l \bar{l}}(L,x) u_{\bar{l}\sigma}(0,{\overrightarrow{0}})     \, ,
\end{equation}
where the coefficient functions on the right are evaluated at the origin $x$ = 0. The Lorentz transformation $L$ transforms the rest momentum $\{\overrightarrow{0},m\}$ to the momentum $p^{\mu}$ = $\{ \overrightarrow{p},p^{t} \}.$ The quantity $M$ = $M(x)$ is a second rank tensor field that is arbitrary until the motion of the particle is considered. 

The displacement function $\epsilon(\Lambda,x,b)$ is also constrained by Eq. (\ref{psi+}), (\ref{Da+}) and (\ref{Dpsi}).  One has
\begin{equation} \label{epsilon1}
 \epsilon^{\mu}(\Lambda,x,b) =  \epsilon^{\mu}(\Lambda) - \Lambda^{\mu}_{\sigma} [M(x)]^{\sigma}_{\nu} x^{\nu} + [M(\Lambda x+b)]^{\mu}_{\nu} (\Lambda x+b)^{\nu} \, ,
\end{equation}
where $M(x)$ is the arbitrary second order tensor field that also appears in (\ref{u}). We drop the first term, $\epsilon^{\mu}(\Lambda)$ = 0. The displacement $\epsilon$ is equal to $b$ when $M$ = 1, i.e. when $M$ is the delta function, the displacements of the field and operators are equal.

The classical motion of a particle follows paths of extreme quantum phase. The relevant phase is the phase of the quantum field, $\delta \Theta$ = $p \cdot M \delta x.$ One finds that $\delta \Theta$ is extreme when $M \delta x$ is parallel to the momentum $p.$ One may write this as $p^{\alpha} = m M^{\alpha}_{\mu} d{ x}^{\mu}/d\tau$ where the quantity $\tau$ turns out to be the proper time. Since the momentum is related to mass by
 \begin{equation} \label{m2} 
\eta_{\alpha \beta} p^{\alpha}p^{\beta} = -m^{2}\, ,
\end{equation}
 one finds immediately from $p^{\alpha} = m M^{\alpha}_{\mu} d{ x}^{\mu}/d\tau$ that
\begin{equation} \label{g2}  g_{\mu \nu} \frac{ dx^{\mu}}{d\tau}\frac{ dx^{\nu}}{d\tau} = -1    \, ,
\end{equation}
where the `curved spacetime metric' $g_{\mu \nu}$ is defined by
\begin{equation} \label{g3}  g_{\mu \nu} \equiv  \eta_{\alpha \beta} M^{\alpha}_{\mu}M^{\beta}_{\nu} \, .
\end{equation}
Further calculations show that the extreme path is the path from general relativity for the curved spacetime metric $g_{\mu \nu}$ (Shurtleff 2008). Since the tensor $M$ is arbitrary, it follows that the metric $g$ is arbitrary and the trajectory can be described in any coordinate system limited only by conditions of continuity and differentiability that are not considered here.

By (\ref{g3}), the second rank tensor field $M(x)$ is a kind of square root field of the curved spacetime metric $g_{\mu \nu}(x).$ In a weak gravitational field, to sufficient accuracy,  in terms of the Newtonian potential $\phi,$ the metric $g_{\mu \nu}(x)$ is, see for example Adler et al. (1965), 
\begin{equation} \label{gNEWT} g_{\mu \nu} = {\mathrm{diag}}(g_{xx},g_{xx},g_{xx},g_{tt}) = {\mathrm{diag}}(1 - 2 \phi,1 - 2 \phi,1 - 2 \phi, -1 - 2 \phi) \, .
\end{equation}
Then, by (\ref{g3}), one choice for $M$ is  
\begin{equation} \label{Mnewt} M^{\alpha}_{\mu} = {\mathrm{diag}}(M_{x},M_{x},M_{x},M_{t}) = {\mathrm{diag}}(1 -  \phi,1 -  \phi,1 -  \phi,1 -  \phi,1 +  \phi) \, ,
\end{equation}
where terms of second order in $\phi$ are dropped. In this way the classical motion determines the value of the field $M(x).$

For the Newtonian potential at Earth we have
\begin{equation} \label{phi} \phi = \phi_{\oplus} + \phi_{\infty} =  - \frac{ G M_{\oplus}}{r_{\oplus} c^{2}} - \frac{ G M_{\odot}}{{\mathrm{1 AU}} c^{2}}+\phi_{\infty}\, ,
\end{equation}
where $G$ is the universal gravitational constant, $M_{\oplus}$ and $M_{\odot}$ are the masses of the Earth and Sun.  Also, $r_{\oplus}$ is the radius of the Earth,  $c$ is the speed of light and $\phi_{\infty}$ is the potential far from the Solar System and far from any other massive object. Note that the $\phi$s have been made unitless by dividing by $c^{2}.$  

In order to equate the potential $\phi$ with $\phi_{\oplus}$ in the expression (\ref{gNEWT}) for the curved spacetime metric $g_{\mu \nu},$ it is necessary to make the potential $\phi_{\infty}$ vanish. To do this one can insist that the trajectory be described in a comoving coordinate system in free fall with the Solar System. In any such coordinate system, the time-time component of the curved spacetime metric is determined, $g_{tt}$ = $-1,$ (Weinberg 1972). In weak fields this component depends on the Newtonian potential $g_{tt}$ = $-1-2\phi_{\infty}.$ Thus in a comoving coordinate system in free fall with the Solar System and at a large distance from the Solar System, $\phi_{\infty}$ = 0. Then the potential $\phi_{\infty}$ of the Galactic gravitational field and more distant masses is canceled out, leaving the local gravitational potential to dominate near the Solar System. Thus Eq. (\ref{gNEWT}) applies with $\phi$ = $\phi_{\oplus}$ in the comoving coordinate system in free fall with the Solar System.

These considerations depend on the phase of the quantum field which, for a given momentum $p,$ is the phase $p \cdot Mx$ of the associated coefficient function $u_{l\sigma}(x ,{\overrightarrow{p}}).$ What about the phase of the particle state with momentum $p?$

First of all, note that the operator $a_{{\sigma}}({\overrightarrow{\Lambda p}})$ removes the same state no matter where the field $\psi^{+}(x)$ is defined, so the operator is independent of $x.$ Since the operator does not communicate location information between the field and the particle states, one cannot be sure that the annihilation operator $a_{\sigma}({\overrightarrow{p}})$ removes a state of the particle with momentum $\{ \overrightarrow{p},p^{t} \}$ in the same reference frame as the quantum field is defined in.  Therefore, one can allow the particle states to be defined in a reference frame with coordinates $y$ while the field is defined in another frame at events with coordinates $x.$ It follows that $x$ = $\lambda y,$ for some Lorentz transformation $\lambda.$ 

The transformation $\{\Lambda,\epsilon\}$ changes the initial phase $p_{0} \cdot y_{0}$ of the particle state with initial momentum $p_{0}$ as follows:
  \begin{equation} \label{phase1}  e^{i p_{0} \cdot y_{0}} \rightarrow e^{i p \cdot (y +\epsilon)} =  e^{i p \cdot y}e^{i p \cdot(- M + \Lambda^{-1} M \Lambda)\lambda y} e^{i p \cdot M_{0} \delta x}\, ,
\end{equation}
where the last expression follows from (\ref{epsilon1}). Also, $p$ = $\Lambda p_{0},$ $y$ = $\Lambda y_{0},$ $M_{0}$ is the tensor in the initial reference frame and $M$ is in the transformed frame with $M_{0}$ = $\Lambda^{-1} M \Lambda $ and $M$ = $\Lambda M_{0} \Lambda^{-1}.$

For a particle moving in a weak gravitational field, one can treat $M$ as constant over fairly large regions, large on the scale of the relevant portion of the quantum field. And on such a scale, the change of the momentum $p$ is often small due to the weak gravitational force. Thus, in many cases, we can treat both $M$ and $p$ as constants on a scale much larger than the scale of the quantum field. 

Then the coefficient of $y$ in the phase of the exponential in (\ref{phase1}) does not depend on $y$ and we have a plane wave in a region where $M$ and $p$ can be considered constant. Defined as a constant times the rate of change of phase with distance, the momentum of the particle state's plane wave is not $p,$ but an `effective momentum' $\bar{p}$ given in
$$ \bar{p} \cdot y = p \cdot y + p \cdot(- M + \Lambda^{-1} M \Lambda) \lambda y = p(1 - M \lambda + \Lambda^{-1} M \Lambda \lambda)\cdot y \, ,  $$
from which it follows that
\begin{equation} \label{pbar}  \bar{p} = p(1 - M \lambda + \Lambda^{-1} M \Lambda \lambda)
\end{equation}
The two momenta $\bar{p}$  and $p$ are equal when $\Lambda$ = 1 or $M$ = 1.

Since the momentum of the particle state plane wave depends on the Lorentz transformations $\Lambda$ and $\lambda,$ it is convenient to work in a fixed reference frame. Appropriate transformations can be applied as needed to go to other frames. A suitable fixed frame is provided by the distribution of Cosmic Microwave Background (CMB) radiation. The observed dipole anisotropy of the CMB implies the Solar System is moving at a speed of 370 km/s = 0.00123$c$ (Fixsen et al. 1996, Lineweaver et al. 1996) with respect to the distribution of CMB radiation. We take the given frame to be the CMB reference frame and, by ignoring the 370 km/s speed compared to the speed of light, we consider the Earth and Sun to be at rest in the CMB frame.

To determine $\Lambda,$ we need to specify an initial reference frame that transforms to the CMB frame with $\Lambda.$ We assume that the initial frame is the particle's rest frame and that the transformation $\Lambda$ takes the rest momentum $\{\overrightarrow{0},m\}$ to the trajectory momentum $p^{\mu}$ = $\{ \overrightarrow{p},p^{t} \},$ i.e. $\Lambda$ is the transformation denoted $L$ in (\ref{u}). We write $\Lambda$ = $L$ in the form
$$ \Lambda^{i}_{k}(p) = L^{i}_{k}(p) = \delta^{i}_{k} + (1+\gamma)^{-1} m^{-2} p^{i}p^{k}\, ,$$
 \begin{equation} \label{L} \Lambda^{i}_{4} = L^{i}_{4} = L^{4}_{i} = m^{-1}p^{i} \quad {\mathrm{and}} \quad \Lambda^{4}_{4} = L^{4}_{4} = \gamma = m^{-1} p^{4}\, ,
\end{equation}
where $i,k \in$ $\{1,2,3\}$ = $\{x,y,z\}$ and $m$ is the mass of the particle. 

The last thing to choose is $\lambda,$ the Lorentz transformation taking the quantum field's frame with coordinates $x$ to the coordinates $y$ for the particle state, $x$ = $\lambda y.$ Remarkably, $\lambda$ = 1 doesn't work. In order for the 3-momentum $\overrightarrow{\bar{p}}$ to increase when the energy ${\bar{p}}^{t}$ increases, one can choose $\lambda$ to be the time inversion $\lambda$ = diag$\{1,1,1,-1\}.$ 

With $M,$ $\Lambda,$ and $\lambda$ determined, one obtains an expression for $\bar{p}.$ By (\ref{Mnewt}), (\ref{pbar}), and (\ref{L}) one finds that
\begin{equation} \label{pbar3} \bar{p}^{\mu} = p^{\mu} (1 +4\phi- 4 \gamma^2 \phi) \approx p^{\mu} (1 - 4 \gamma^2 \phi) \, ,
\end{equation}
where $\phi$ is small, i.e. $\mid \phi \mid \ll 1,$  $\bar{p}^{k}$ indicates the spatial part of the effective four-momentum, and $k \in$ $\{1,2,3\}.$ Also, $\bar{p}^{\,t}$ = $\bar{p}^{\,4}$ = $\bar{E}+mc^2$ is the effective total energy of the particle state, including the rest energy. 

To transform spectra from one gravitational potential to another, we need the time component of the momentum in (\ref{pbar3}). One finds that
\begin{equation} \label{E1} E  = E_{0} - 4 \phi \frac{  (E_{0} + mc^{2})^{3}}{m^{2}c^{4}}  \, ,
\end{equation}
where the total particle state energy is $E + mc^{2}$ = $\bar{p}^{\,t}$ and $E_{0}$ is the kinetic energy at $\phi$ = 0. The term $mc^{2}$ has been cancelled from both sides of the equation. The cubic term gets one ${p}^{t}$ = ${E_{0}}+mc^2$ from the factor $p$ preceding the parentheses in (\ref{pbar}) and (\ref{pbar3}) while the other two come from $\gamma$ = $p^{t}/m.$ 

For the energy cubed term to be important, the energy $E_{0}$ at $\phi$ = 0 must be much larger than the rest energy $mc^{2},$ since the potential $\phi$ is small in weak gravitational fields. In that case, (\ref{E1}) reduces to (\ref{Eintro}).  

It may be that some massive particles have $M,$ $\Lambda,$ and $\lambda$ determined as above to obtain (\ref{E1}), while other particles are described with different assumptions for $M,$ $\Lambda,$ and $\lambda.$ As mentioned in the Introduction, accelerator experiments have shown that electrons do not obey (\ref{E1}) at least with $\phi$ = $\phi_{\oplus}.$ It may be that the comoving coordinate system for electrons occurs at so small a scale that the potential due to the Sun and Earth vanishes, $\phi_{\oplus}$ = 0, just as the potential due to the Galaxy and larger structures vanishes, $\phi_{\infty}$ = 0, in a comoving coordinate system in free fall with the Solar System. Or it could be that $\lambda$ is the identity and $M_{t}$ in (\ref{Mnewt}) is $M_{t}$ = $-1-\phi$ with $M_{x}$ and $\Lambda$ as above. These choices for $M,$ $\Lambda,$ and $\lambda$ would give the same curved spacetime metric (\ref{gNEWT}) as above, but (\ref{pbar}) now yields $\bar{p}^{\,4}$ = $p^{4} + O(\phi^{2})$ instead of (\ref{E1}), meaning the energy of the particle state would not depend linearly on the gravitational potential $\phi.$ 

In this paper only protons are considered as cosmic ray primaries and protons are assumed to have potential-dependent particle state energies as obtained above for (\ref{E1}).

By the definition of `weak field', as an energetic particle $E_{0} >> mc^{2}$ moves in a weak field, its trajectory is accelerated by only a small amount. Thus the energy $E_{0}$ is very nearly equal to the kinetic energy all along the particle's path in weak gravitational fields. Hence, $E_{0}$ approximates the `trajectory energy'.

Consider a particle entering a nonzero gravitational potential $\phi < 0.$ Its particle state energy $E$ increases by (\ref{Eintro}), while its trajectory energy increases by a very small amount. Such a particle then has two kinds of energy: the particle state energy and the trajectory energy. And these can differ significantly at sufficiently high energies. Measurements of the particle's trajectory by time-of-flight, say, should produce the energy $E_{0},$ but, in a particle-state-changing collision, the particle should deposit its particle state energy $E.$

\section{CRs at Earth to VHE Gamma Rays at Earth via the SNR}

Given the observed cosmic ray spectrum and the CR's energy dependence on gravitational potential in (\ref{Eintro}), one can deduce the CR spectrum at a typical SNR. Applying astroparticle physics to the collisions of CRs and ambient matter, one can deduce the gamma ray spectrum produced by the collisions at the SNR. The VHE gamma ray spectrum at the SNR and the assumption that gamma rays also obey a modified version of (\ref{Eintro}) allows one to deduce the expected VHE gamma ray spectrum at Earth. The process is the topic of this section.

The observed CR flux $dN_{p}/dEp_{\oplus}$ (Afanasiev (Yakutsk) 1996, AGASA 2003, Auger 2007, Bird (HiRes) 1994,  Cronin et al. 1997, Grigorov et al. (Proton) 1991, Lawrence et al. (Haverah) 1991, Nagano et al. (Akeno 1) 1984,  Nagano et al. (Akeno 2) 1992, Seo et al. (LEAP) 1991) is sectioned by energy and each section is fit. The highest energy CRs observed are not relevant here, so the data has not been updated for the latest values. Computer software (Mathematica 6.0) is used to determine best fit values of $\{a_{i},b_{i},c_{i}\}$ in $\log{(dN_{p}/dEp_{\oplus})}$ = $a_{i} + b_{i} \log{(Ep_{\oplus})} + c_{i}[\log{(Ep_{\oplus})}]^{2}$ for five intervals of $\log{(Ep_{\oplus})}:$ $\{\log{(Ep_{\oplus})}_{\mathrm{initial}},$ $\log{(Ep_{\oplus})}_{\mathrm{final}}\}$ = $\{8.33,10.64\},$ $\{10.64,14.92\},$ $\{14.92,16.68\},$ $\{16.68,18.54\},$ $\{18.54,20.5\},$ where $Ep_{\oplus}$ has units of eV. A collection of points $\{\log{Ep_{\oplus}},\log{dN_{p}/dEp_{\oplus}}\},$ calculated every $\delta \log{Ep_{\oplus}}$ = 0.1 along the five fits, is made from  $\log{(Ep_{\oplus})}$ = $8.0$ to $22.0.$ Minor adjustments are made to smooth the transition from one section to the next. A third-order interpolation is constructed based on the collection of points. The data and interpolating function are shown in Fig. \ref{CRearth}.

Eq. (\ref{E1}) from Sec. \ref{PDPSE} relates the CR energy $Ep_{\oplus}$ at Earth ($\oplus$) and the CR energy $Ep_{S}$ at the supernova remnant ($S$). However, the expressions are messy, so, for the purpose of displaying the intermediate results in this section, the simpler and only slightly less accurate formula (\ref{Eintro}) is used. 

\begin{figure}[ht]  
\centering
\vspace{0cm}
\hspace{0in}\includegraphics[0,0][360,360]{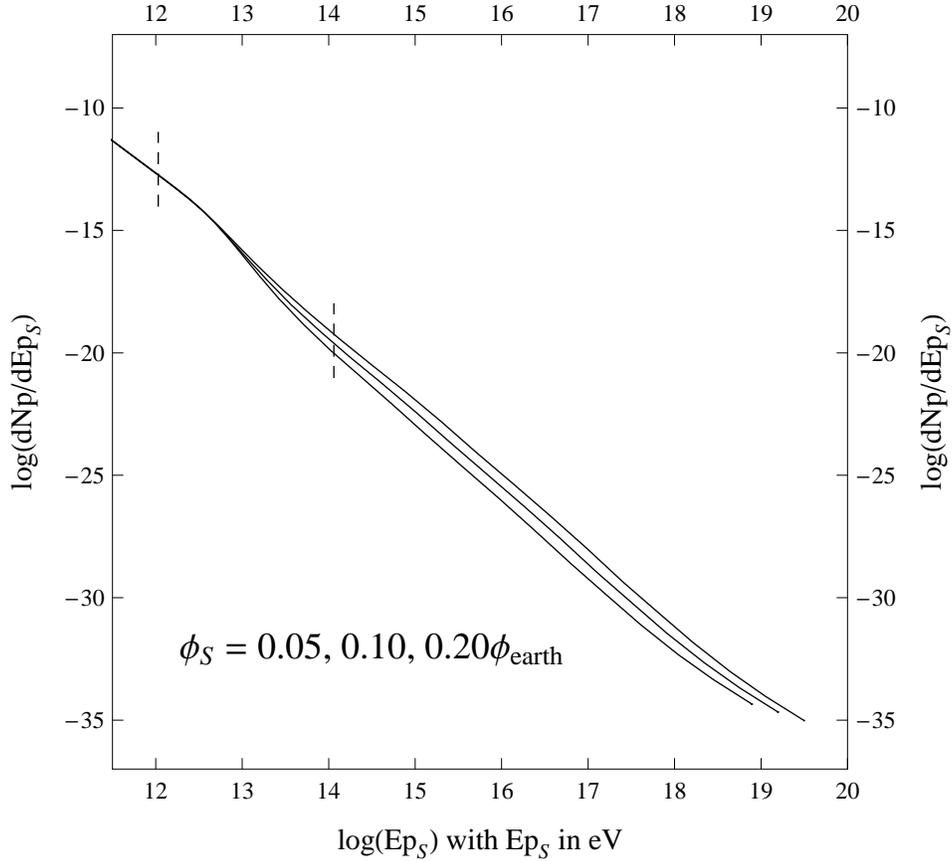}
\caption{{\it{The Expected Cosmic Ray Spectrum at a Typical SNR.}}  The spectrum bends in the region corresponding to the VHE gamma ray spectrum. Compare this with the same region of the CR spectrum at Earth, see Fig. \ref{CRearth}. The bend in the CR flux at the SNR reflects the CR proton energy dependence on gravitational potential, (\ref{dNpdEpSNR}). The displayed curves are for potentials $\phi_{S}$ differing by successive factors of 2. By formula (\ref{Eintro}), the CR energy at Earth is $Ep_{\oplus}$ = $Ep_{0} - 4 \phi_{\oplus} Ep_{0}^{3}/mp^{2}$ and at the SNR, $Ep_{S}$ = $Ep_{0} - 4 \phi_{S} Ep_{0}^{3}/mp^{2},$ so at high energies, we have $Ep_{S}$ = $Ep_{\oplus} \phi_{S}/\phi_{\oplus}.$ It follows that, at high energies, a given CR flux appears at $Ep_{S}$ values separated by about $\Delta (\log{Ep_{S})}$ = $\log{2}$ = 0.30. }
\label{CRatSNR}
\end{figure}

\begin{figure}[ht] 
\centering
\vspace{0cm}
\hspace{0in}\includegraphics[0,0][360,360]{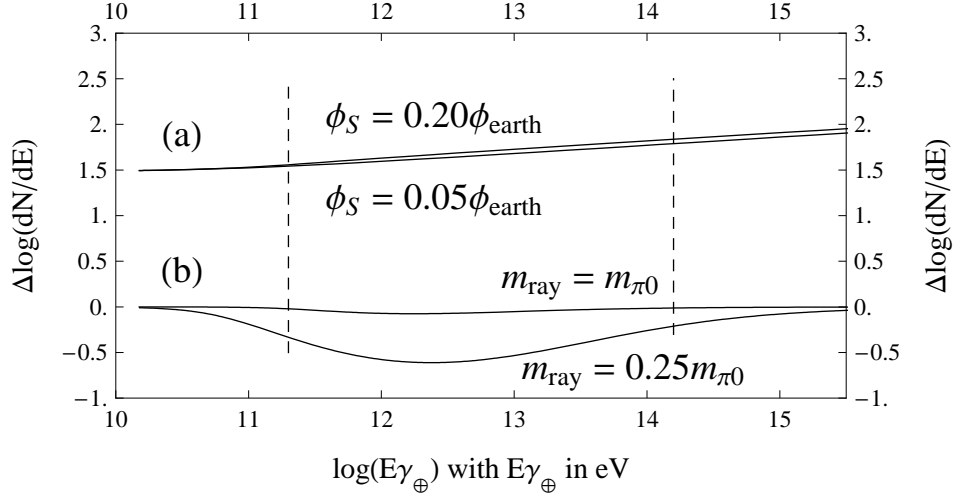}
\caption{{\it{Contributions to the VHE Gamma Ray Spectrum from the $p$-$p$ Cross Section and the Energy Transitions of the CR and Gamma Ray Fluxes .}}  [A modified $\delta$-functional approach is used for this rough sketch: instead of $E\gamma_{S}$ = $0.17Ep_{S},$ we employ $E\gamma_{0}$ = $0.17Ep_{0}.$] (a) $\log{\sigma_{inel}}$ for $m_{ray}$ = $0.5m_{\pi^{0}}.$ The log of the inelastic $p$-$p$ cross section slopes steadily upward in the region corresponding to the observed VHE gamma ray spectra. (b) $\log{\left(dEp_{\oplus}/dEp_{S}\right)}$ + $\log{\left(dE\gamma_{S}/dE\gamma_{\oplus}\right)}$ for $\phi_{S}$ = $0.1\phi_{\oplus}.$ Relating the CR flux at Earth to the flux at the SNR and relating the VHE gamma ray flux at the SNR to the flux at Earth introduces the factors  $(1-12 \phi_{\oplus}\, \frac{Ep_{0}^{2}}{m_{p}^{2}c^{4}})/(1-12 \phi_{S} \, \frac{Ep_{0}^{2}}{m_{p}^{2}c^{4}})$ and $(1-12 \phi_{S}\, \frac{E\gamma_{0}^{2}}{m_{ray}^{2}c^{4}})/(1-12 \phi_{\oplus} \, \frac{E\gamma_{0}^{2}}{m_{ray}^{2}c^{4}}),$ respectively.  See (\ref{dNpdEpSNR}) and (\ref{dNgdEgEarth}). At both high and low energies the product of the factors is unity, but in the region of the VHE gamma ray spectrum the log of the product of the factors changes by up to 0.5 (= $\log{3}$) depending on $m_{ray}$ and $\phi_{S}.$}
 \label{sigmadEdE}
\end{figure}

The flux at the SNR, $dN_{p}/dEp_{S},$ is related to the flux at earth by 
\begin{equation} \label{dNpdEpSNR}  \frac{dN_{p}}{dEp_{S}} = \frac{dN_{p}}{dEp_{\oplus}} \, \frac{\frac{dEp_{\oplus}}{dEp_{0}}}{\frac{dEp_{S}}{dEp_{0}}} = \frac{dN_{p}}{dEp_{\oplus}} \, \frac{1-12 \phi_{\oplus}\, \frac{Ep_{0}^{2}}{m_{p}^{2}c^{4}}}{1-12 \phi_{S} \, \frac{Ep_{0}^{2}}{m_{p}^{2}c^{4}}}\, ,
\end{equation}
where the CR energy at null potential $Ep_{0}$ gives both the CR energy at Earth, $Ep_{\oplus} $ = $Ep_{0} - 4 \phi_{\oplus} Ep_{0}^{3}/m_{p}^{2}c^{4},$ and the energy of the same CR at the SNR, $Ep_{S} $ = $Ep_{0} - 4 \phi_{S} Ep_{0}^{3}/m_{p}^{2}c^{4}.$ Many SNRs contribute to the observed CR spectrum at Earth, so the SNR is called a `typical' SNR. Fig. \ref{CRatSNR} shows the CR flux at the typical SNR using various values of $\phi_{S}.$ The absolute normalization is meaningless because it depends on the unknown number of contributing SNRs and the unknown distances to the SNRs.

\begin{figure}[ht]  
\centering
\vspace{0cm}
\hspace{0in}\includegraphics[0,0][360,360]{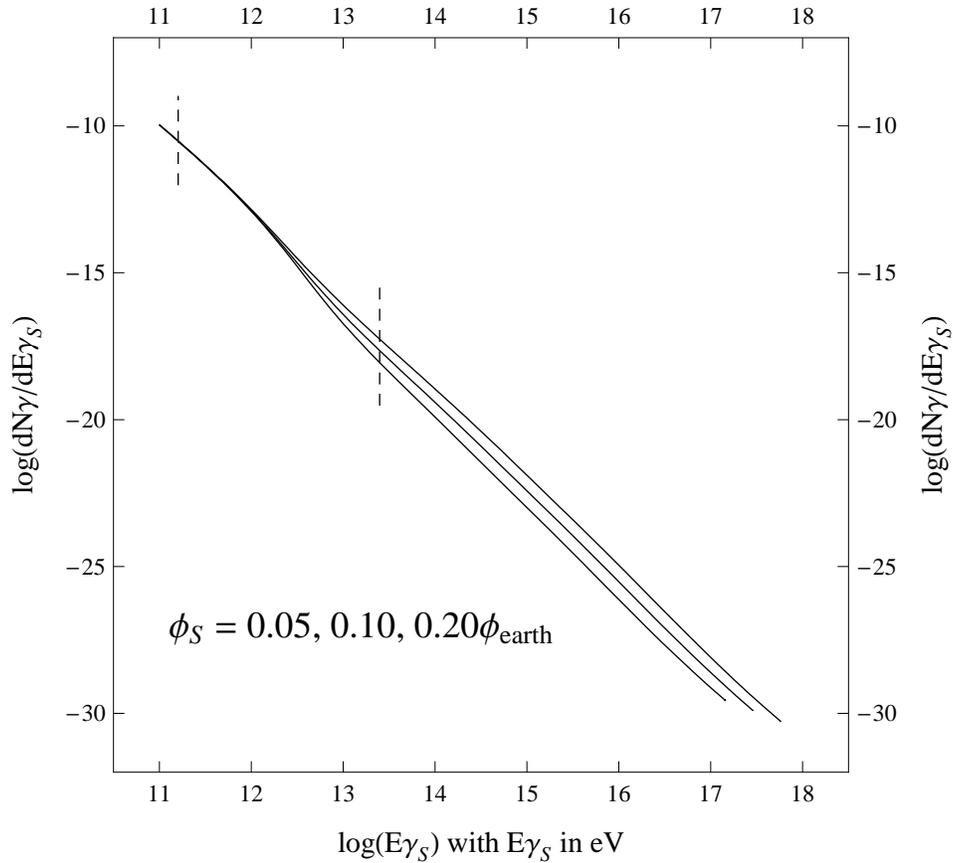}
\caption{{\it{The Expected VHE Gamma Ray Spectrum at a Typical SNR.}} CR protons must have more energy than the secondary gamma rays they produce. So the gamma ray spectra are calculated by integrating over those CR protons that have more energy. The calculation follows the method of Kelner et al. (2006). Note how the gamma ray spectrum at the SNR in this graph mirrors the CR spectrum at the SNR in Fig. \ref{CRatSNR}, except that the gamma ray energies are lower than the CR energies. }
\label{VHEatSNR}
\end{figure}

It is assumed that VHE gamma rays are produced when CR protons strike ambient protons at the SNR. The VHE gamma ray flux at the SNR, $dN_{\gamma}/dE\gamma_{S},$ is determined from the CR flux at the SNR, $dN_{p}/dEp_{S},$  using an analytical expression for secondary gamma-ray spectra from inelastic proton-proton interactions presented by Kelner et al. (2006). One has
\begin{equation} \label{d2Ngamma} \frac{d^{2}N_{\gamma}}{dE\gamma_{S}\, dEp_{S}} = \frac{d^{2}N_{\gamma}}{dx \, dN_{collision}}\, \frac{dN_{collision}}{dN_{p}} \, \frac{dN_{p}}{dEp_{S}} \, \frac{dx}{dE\gamma_{S}} \, ,
\end{equation}
where $d^{2}N_{\gamma}/dx dN_{collision}$ = $F_{\gamma}(x,E_{p})$ indicates the number of gamma rays in the interval $(x,x+dx)$ per collision and $x$ = $E\gamma_{S}/Ep_{S}.$ Also, $dN_{collision}/dN_{p}$ = $c n_{H} \sigma_{inel}$ where $c$ is the speed of light, $n_{H}$ density of ambient protons, and $\sigma_{inel}$ is the inelastic $p$-$p$ cross section. It follows that
\begin{equation} \label{VHEfluxS}
\frac{dN_{\gamma}}{dE\gamma_{S}} = \int_{E\gamma_{S}}^{\infty} F_{\gamma}\left(\frac{E\gamma_{S}}{Ep_{S}},Ep_{S}\right) \, \sigma_{inel} \left( Ep_{S} \right) \, \frac{dN_{p}}{dEp_{S}} \, \frac{dEp_{S}}{Ep_{S}} \, ,
\end{equation}
where the absolute normalization $A_{\gamma}$ is selected to simplify the expression by adjusting the units and canceling the factor $c n_{H},$ $A_{\gamma}c n_{H}$ = 1. The CR spectrum at the SNR, $dN_{p}/dEp_{S},$ is given by (\ref{dNpdEpSNR}) and expressions for $F_{\gamma}$ and $\sigma_{inel}$ are given by Kelner et al. (2006). The approach by Kelner et al. (2006) introduces a lower limit of $10^{11}$ eV, so $E\gamma_{S} >$ 0.1 TeV. The VHE gamma ray spectra for various values of the gravitational potential at the SNR $\phi_{S}$ are displayed in Fig. \ref{VHEatSNR}. 

A VHE gamma ray can interact with a variety of particles, low energy photons or low energy massive particles, to produce a massive particle such as a neutral pion or eta, an electron, a positron, etc. And the particle produced has practically the same energy as the VHE gamma ray. Even if this never happens, one can imagine it could happen anywhere along the gamma ray's trajectory. It is inferred that the energy of the gamma ray would have the same dependence on gravitational potential as any massive particle it could produce. Since it could produce many different massive particles, let us try using the massive particle Eq. (\ref{Eintro}) with the mass $m$ in (\ref{Eintro}) allowed to be some average mass, an average of the masses of a neutral pion or eta, electron, positron, etc. 

In effect, (\ref{Eintro}) is applied to VHE gamma rays with the mass $m \rightarrow$ $m_{ray}$ allowed to vary freely. A best-fit value for $m_{ray}$ is obtained by comparing the observed VHE spectra with the VHE spectrum predicted for a typical SNR. None of this implies that a gamma ray has mass or that the imagined interactions actually occur.

\begin{figure}[ht] 
\centering
\vspace{0cm}
\hspace{0in}\includegraphics[0,0][360,360]{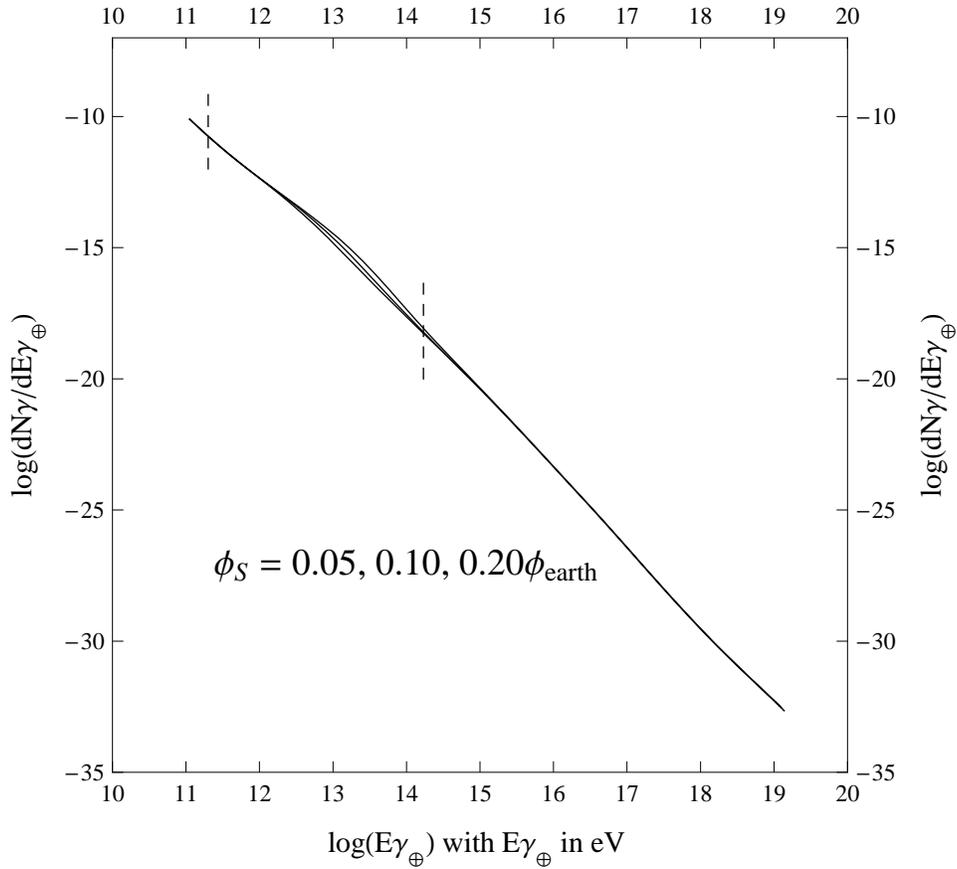}
\caption{{\it{The Expected VHE Gamma Ray Spectrum at Earth; Dependence on the SNR Potential $\phi_{S}.$}} At high energies, just as expected from Fig. \ref{sigmadEdE}b, taking the CR spectrum at Earth backwards to the SNR cancels taking the VHE gamma ray spectrum at the SNR forward to Earth, as far as dependence on $\phi_{S}$ is concerned. The only dependence on $\phi_{S}$ is in the region of interest, the region of the feature in the observed VHE gamma ray spectra.   }
 \label{VHEatEarth}
\end{figure}

\begin{figure}[ht]  
\centering
\vspace{0cm}
\hspace{0in}\includegraphics[0,0][360,360]{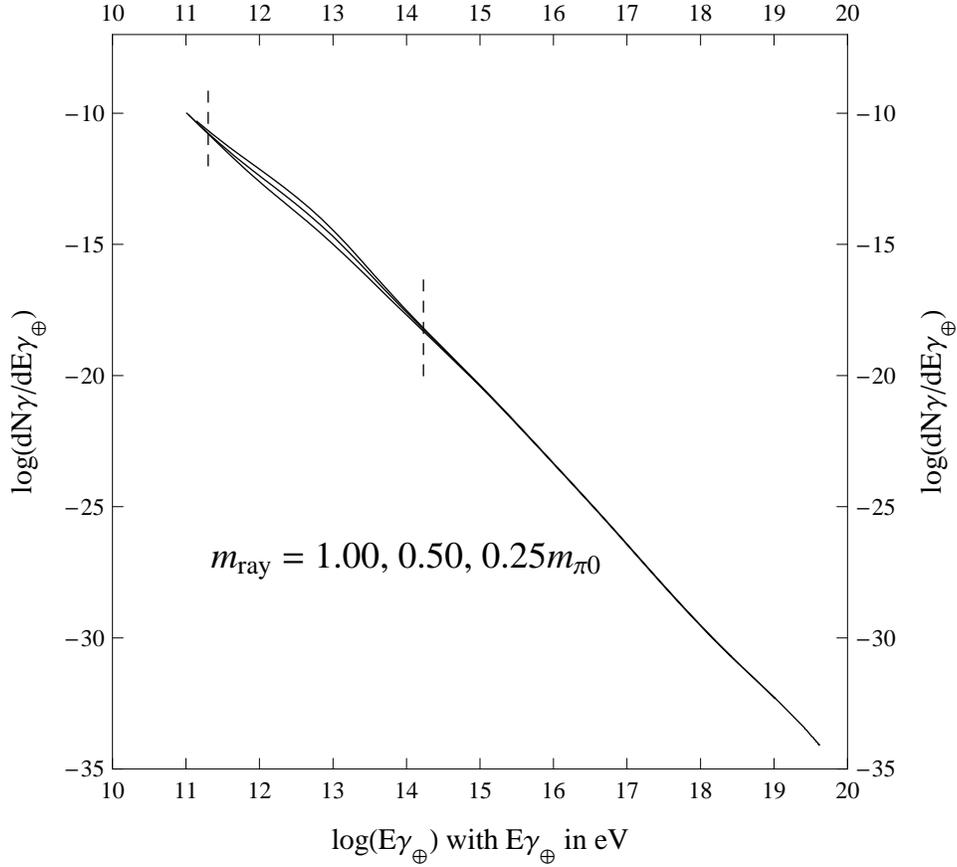}
\caption{{\it{The Expected VHE Gamma Ray Spectrum at Earth; Dependence on Mass $m_{ray}.$}} Just as with the dependence on the SNR potential $\phi_{S},$ the dependence on the quantity $m_{ray}$ is confined to the region of interest. The reason is the same, in the limit of large or small values of $m_{ray},$ the factor whose log is plotted in Fig. \ref{sigmadEdE}b approaches unity. The curves are drawn left to right in the order that the $m_{ray}$s are listed.  }
\label{VHEatEarthMASS}
\end{figure}

Then applying (\ref{Eintro}) with $m \rightarrow$ $m_{ray}$ to the VHE gamma ray flux $dN_{\gamma}/dE\gamma_{S}$ at a typical SNR in (\ref{VHEfluxS}) yields the  expected VHE gamma ray flux $dN_{\gamma}/dE\gamma_{\oplus}$ at Earth received from a typical SNR. One finds
\begin{equation} \label{dNgdEgEarth}  \frac{dN_{\gamma}}{dE\gamma_{\oplus}} = \frac{dN_{\gamma}}{dE\gamma_{S}} \, \frac{\frac{dE\gamma_{S}}{dE\gamma_{0}}}{ \frac{dE\gamma_{\oplus}}{dE\gamma_{0}}} = \frac{dN_{\gamma}}{dE\gamma_{S}} \, \frac{1-12 \phi_{S}\, \frac{E\gamma_{0}^{2}}{m_{ray}^{2}c^{4}}}{1-12 \phi_{\oplus} \, \frac{E\gamma_{0}^{2}}{m_{ray}^{2}c^{4}}}\, ,
\end{equation}
where the energy at null potential $E\gamma_{0}$ is a parameter that gives both the gamma ray energy at the Earth $E\gamma_{\oplus} $ = $E\gamma_{0} - 4 \phi_{\oplus} E\gamma_{0}^{3}/m_{ray}^{2}c^{4}$ and the energy of the same gamma ray at the SNR $E\gamma_{S}$ = $E\gamma_{0} - 4 \phi_{S} E\gamma_{0}^{3}/m_{ray}^{2}c^{4}.$ Figs. \ref{VHEatEarth} and \ref{VHEatEarthMASS} gives the VHE gamma ray spectra at Earth for various values of $\phi_{S}$ and $m_{ray}.$

The process described in this section produces VHE gamma ray fluxes and energies from the observed CR spectrum at Earth given the two parameters $\phi_{S}$ and $m_{ray}.$ In the next section the values of $\phi_{S}$ and $m_{ray}$ are found that give a best fit to the observed VHE gamma ray fluxes.   

\section{Fitting the Observed VHE Gamma Ray Spectra}

Galactic CR protons are deflected as they travel in the Milky Way and directional information is lost, whereas a gamma ray trajectory aligns with the other electromagnetic emissions from its source. One can see where gamma rays come from. Since the CR spectrum at Earth represents many sources, it is appropriate to compare the VHE gamma ray spectra found in Sec. 3 with a VHE gamma ray spectrum obtained by combining many spectra. 

The VHE gamma ray spectra found in Sec. 3 depend on the SNR gravitational potential $\phi_{S}$ and the mass $m_{ray}$ needed to apply (\ref{Eintro}) to gamma rays. Given a value for $\phi_{S}$ and and a value for $m_{ray}$ the process in Sec. 3  determines a proposed typical VHE gamma ray spectrum based on the CR spectrum observed on Earth. Call the VHE gamma ray spectrum obtained the `proposed spectrum'; it is a function of $\phi_{S}$ and $m_{ray}.$ 

Since VHE gamma rays can also be produced in the lepton scenario, it is important to select spectra produced by the hadron scenario exclusively, with little contamination by VHE gamma rays produced by the lepton scenario. Among the qualifying sources are the spectra from the SNRs RX J1713.7-3946 ($h_{1}$ = 28 data points) and RX J0852.0-4622 ($h_{2}$ = 14 data points) and from the Galactic Center ridge ($h_{3}$ = 9 data points)  (Aharonian et al. 2007a, 2007b, 2006). In this section, these 51 data points are combined and compared with the VHE gamma ray spectra found in Sec. 3.

To compare spectra, the absolute normalization of the proposed spectrum is not changed while the absolute normalizations of the observed VHE gamma spectra are adjusted to minimize the separation of the experimental flux points to the proposed spectrum. Let $Flux(E_{\gamma},\phi_{S},m_{ray})$ =  $dN_{\gamma}/dE\gamma_{\oplus}$ be the proposed flux and let $Flux_{ij}^{0}(E_{ij})$ =  $(dN_{\gamma}/dE_{\gamma})_{ij}$ be the $j$th experimental flux measured at an energy of $E_{ij}$ from the $i$th source.  Then, for each of the three sources, the value of $\Delta_{i},$ $i \in\{1,2,3\},$ is found that minimizes the average separation $\bar{\delta}_{i}(\Delta_{i})$ given by
\begin{equation} \label{findDelta}
\bar{\delta}_{i}(\Delta_{i}) = \left(\frac{1}{h_{i}}\sum_{j}{\left[\Delta_{i} + \log{\left(Flux_{ij}^{0}(E_{ij})\right)} - \log{\left(Flux(E_{ij},\phi_{S},m_{ray})\right)}\right]^{2}}\right)^{1/2} \, ,
\end{equation}
where $h_{i}$ is the number of data points for the $i$th source. The normalization-adjusted flux at an energy of $E_{ij}$ is given by
\begin{equation} \label{MovedFlux}
\log{\left(Flux_{ij}(E_{ij})\right)}= \Delta_{i} + \log{\left(Flux_{ij}^{0}(E_{ij})\right)} \, .
\end{equation}
The collection of the three adjusted spectra are now to be compared with the proposed spectra from Sec. 3.

Once the absolute normalizations of the three sources are adjusted, the goodness-of-fit of a proposed VHE gamma ray spectrum is calculated. The goodness-of-fit function $g(\phi_{S},m_{ray})$ is the average separation of the proposed spectrum from the $\sum h_{i} -1 $ = 50 observed flux data points. The rejected data point is the flux with a gamma ray energy of 169.79 TeV in the spectrum of RX J1713.7-3946, rejected because it has low significance (Aharonian et al. 2007a).  The goodness-of-fit function $g(\phi_{S},m_{ray})$ is  defined by
\begin{equation} \label{goodfit}
g(\phi_{S},m_{ray}) \equiv \left(\frac{1}{\sum{h_{i}}-1}\sum_{i,j}{\left[ \log{\left(Flux_{ij}(E_{ij})\right)} - \log{\left(Flux(E_{ij},\phi_{S},m_{ray})\right)}\right]^{2}}\right)^{1/2} \, .
\end{equation}
By varying $\phi_{S}$ and $m_{ray}$ and calculating $g(\phi_{S},m_{ray})$ for each, a best fit is found. The best fit and some close runnersup are given in Table \ref{table:Fits}. 

The best fit has a goodness-of-fit of 0.074 with $\phi_{S}$ = $0.11\phi_{\oplus},$ $m_{ray}$ = $0.45 m_{\pi^{0}},$ where $m_{\pi^{0}}$ is the mass of a neutral pion and $\phi_{\oplus}$ = $-1.06\times 10^{-8}$ is the Newtonian potential at Earth due to the Sun and Earth referenced to a null potential at great distances from the Solar System, see (\ref{phi}).  The best fit proposed VHE gamma ray spectrum and the three normalization-adjusted observed spectra including the point at 169.79 TeV are plotted in Fig. \ref{bestfit}.

Tables \ref{table:CRenergy}, \ref{table:VHEenergy} and \ref{table:Flux} contain sample numerical values for the various quantities discussed in this article. Wherever needed, the tables use the best fit value of the gravitational potential at the SNR, $\phi_{S},$ and the best fit value for $m_{ray}.$

\section{Discussion and Conclusion} \label{discus}

\begin{figure}[ht]  
\centering
\vspace{0cm}
\hspace{0in}\includegraphics[0,0][360,360]{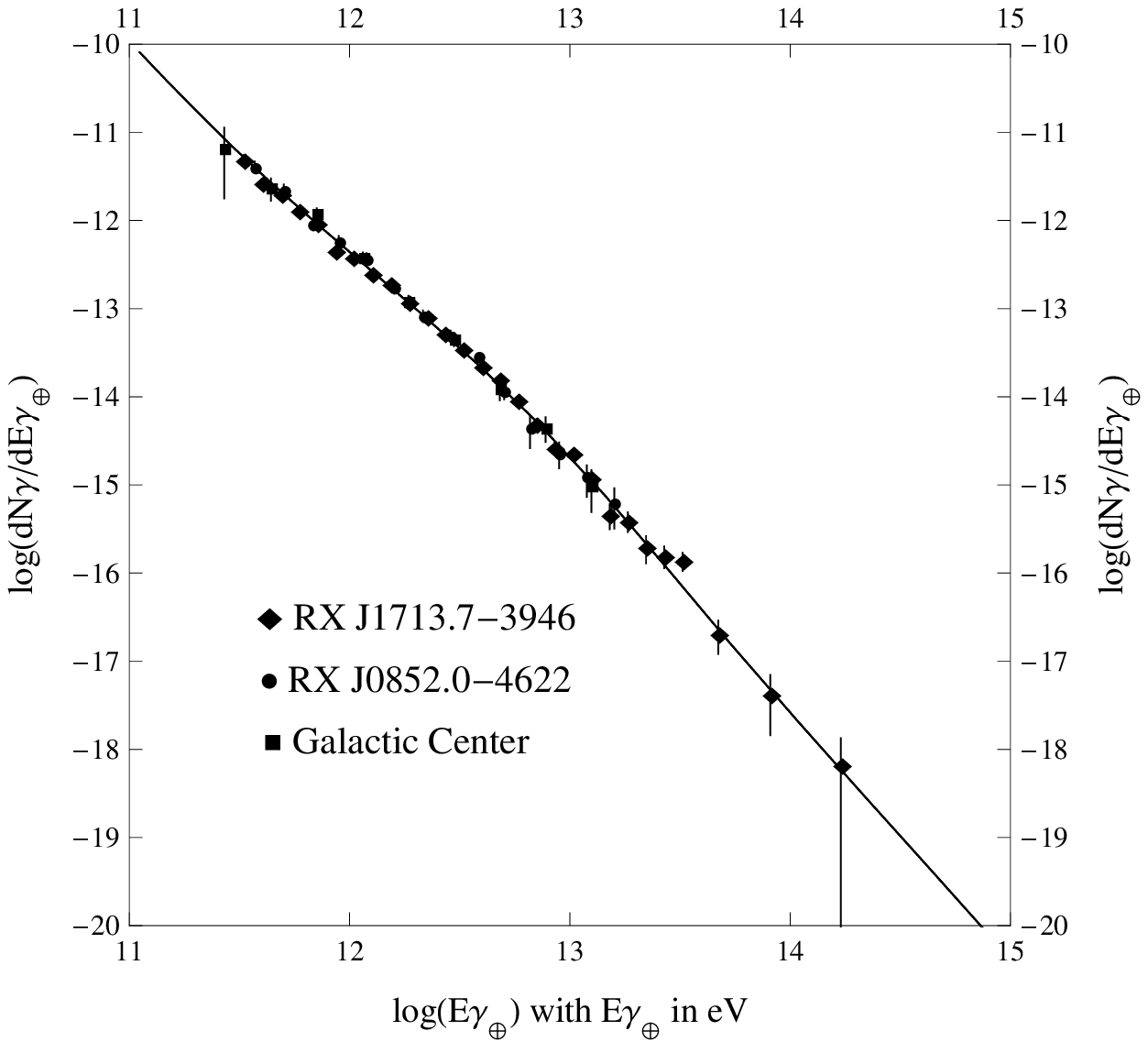}
\caption{{\it{The Best Fit to the Observed VHE Gamma Ray Spectrum at Earth.}} The graph shows the proposed VHE gamma ray spectrum at Earth evaluated in the region of interest for the best fit values in Table \ref{table:Fits}. The data points are compiled from the published spectra of SNRs RX J1713.7-3946  and RX J0852.0-4622 and from the Galactic Center ridge (Aharonian et al. 2007a, 2007b, 2006). The absolute normalizations of the three spectra are adjusted by the values $\Delta_{i}$ listed in Table \ref{table:Fits}, see (\ref{MovedFlux}). The three highest energy data points are outnumbered by the 48 data points of lower energy, so fitting the three highest energy data points so closely is remarkable. }
\label{bestfit}
\end{figure}

The explanation of the VHE gamma ray spectra of typical SNRs presented in this paper is based on the dependence of particle state energy on gravitational potential as displayed in (\ref{Eintro}). A proton or a gamma ray in the near-null potential at an SNR has less energy than it has in the gravitational potential at the Earth. The quality of the fit to the data in Fig. \ref{bestfit} supports this explanation.

The quality of the fit supports the application of formula (\ref{Eintro}) to photons even though photons are massless and (\ref{Eintro}) applies to massive particles. It is argued that the photon should have particle states with energies almost the same as the energies of various massive particle states. The coincidence of photon particle state energy with massive particle state energy is supported by the VHE gamma ray spectra.

For a given source of VHE gamma rays, the distance to the source and the number of CRs each contributes to the CR spectrum at Earth are not knowable to sufficient accuracy to be useful in setting the absolute normalization of the VHE gamma ray spectra. One might include as free parameters the $\Delta_{i}$s needed to adjust the absolute normalizations of the three VHE gamma ray spectra.

One of the two parameters needed to obtain an expected VHE gamma ray spectrum from the observed CR spectrum at Earth is the mass $m_{ray}.$ Since the best fit mass $m_{ray}$ is less than the neutral pion mass, there must be some interaction that allows a VHE gamma ray to transfer almost all its energy to a particle with less mass than the neutral pion. Keep in mind that transferring energy to electrons and positrons does not contribute to $m_{ray}$ because they do not obey (\ref{Eintro}). 

The best fit gravitational potential at the SNR is about ten percent of the potential at the Earth's surface. Using this to imply something about the density of, the dimensions of, and the distance to the clouds of ambient matter that produce the VHE gamma rays is beyond the scope of this article.   

Consider Fig. 7. Draw a straight line on the graph from a point near 0.3 TeV to a point near 30 TeV, on the graph this is from $\log{E\gamma_{\oplus}}$ = 11.5 to 13.5. Note the `bump' in the VHE gamma ray spectra. The same bump is seen in all three observed spectra which implies that some common mechanism is in place to produce the bump. That the same bump can be replicated by the best fit values of $\phi_{S}$ and $m_{ray}$ shows how well the explanation relates the CR spectrum at Earth with the VHE gamma ray spectrum at Earth. 

The tightness of the fit in Fig. \ref{bestfit} suggests that the feature found in the VHE gamma ray spectrum of SNR RX J1713.7-3946 and the other SNRs plotted in Fig. \ref{bestfit} exhibits particle behavior consistent with the simple expression (\ref{Eintro}) of a gravitational potential-dependent relationship between CR and VHE gamma ray energies at Earth and the energies at the SNR.

  \begin{table*}[!ht]
    \caption{There are low goodness-of-fit values $g(\phi_{S},m_{ray})$ for proposed VHE gamma ray spectra with $m_{ray}$ from 0.35 to $0.55m_{\pi^{0}}$ and with $\phi_{S}$ from 0.09 to $0.13 \phi_{\oplus},$ where $m_{ray}$ is the average mass of possible gamma ray produced particles and $\phi_{S}$ is the gravitational potential at the SNR. The constant $\phi_{\oplus}$  is the potential at the Earth's surface due to the Earth and Sun referenced to zero potential at infinite distance, see (\ref{phi}). The $\Delta$s adjust the absolute normalizations of the three observed spectra to each proposed spectrum.  \vspace{0.5cm} }
    \label{table:Fits}
    \centering
    \begin{tabular*}{\textwidth}{@{\extracolsep{\fill}}|c|c|c|c|c|c|c|}
      \hline
      \# &  $g(\phi_{S},m_{ray})$ & $m_{ray}/m_{\pi^{0}}$  & $\phi_{S}/\phi_{\oplus}$   & $\Delta_{1}$  & $\Delta_{2}$  & $\Delta_{3}$ \\\hline\hline
1    &  0.074  &  0.45  &  0.11  &    1.75  &  1.62  &  0.91  \\\hline
2    &  0.076  &  0.35  &  0.11  &    1.67  &  1.54  &  0.83  \\\hline
3    &  0.077  &  0.45  &  0.09  &    1.72  &  1.60  &  0.89  \\\hline
4    &  0.078  &  0.55  &  0.11  &    1.81  &  1.69  &  0.97  \\\hline
5    &  0.078  &  0.45  &  0.13  &    1.78  &  1.65  &  0.93  \\\hline
6    &  0.079  &  0.55  &  0.09  &   1.79  &  1.67  &  0.96  \\\hline
7    &  0.080  &  0.35  &  0.09  &    1.63  &  1.50  &  0.80  \\\hline
8    &  0.080  &  0.35  &  0.13  &   1.71  &  1.57  &  0.85  \\\hline
9    &  0.082  &  0.55  &  0.13  &   1.83  &  1.71  &  0.99  \\\hline
    \end{tabular*}
  \end{table*}

\pagebreak
\clearpage
 \begin{table*}[!h]
    \caption{The particle state energies of 11 CR protons in the gravitational potential of the Earth ($\oplus$), in a null potential ($0$), and in the potential at an SNR ($S$), providing examples of Eq. (\ref{Eintro}) which show how particle state energy depends on gravitational potential. For the potential at the SNR, $\phi_{S},$ the best fit value from Table \ref{table:Fits} is used. \vspace{0.5cm}}
    \label{table:CRenergy}
    \centering
    \begin{tabular*}{0.5\textwidth}{@{\extracolsep{\fill}}|c|c|c|c|}
      \hline
      \# & $Ep_{\oplus}$  (TeV) & $Ep_{0}$  (TeV) & $Ep_{S}$  (TeV)  \\\hline\hline
1. &  1.00 &  0.96 &  0.96    \\\hline
2. &  2.00 &  1.74 &  1.77   \\\hline
3. &  4.00 &  2.87 &  2.99    \\\hline
4. &  8.00 &  4.27 &  4.68    \\\hline
5. &  16.0 &  5.94 &  7.05    \\\hline
6. &  32.0 &  7.94 &  10.6    \\\hline
7. &  64.0 &  10.4 &  16.3   \\\hline
8. &  128. &  13.4 &  26.0  \\\hline
9. &  256. &  17.1 &  43.4    \\\hline
10. &  512. &  21.7 &  75.6    \\\hline
11. &  1024. &  27.5 &  137.    \\\hline
    \end{tabular*}
  \end{table*}

 \begin{table*}[!ht]
    \caption{The energies of 11 VHE gamma rays in the gravitational potential of the Earth ($\oplus$), in a null potential ($0$), and in the potential at an SNR ($S$). While (\ref{Eintro}) only applies to massive particles, it is applied here for massless gamma rays by considering virtual though possible processes that produce massive particles with essentially the same energies as the gamma rays. The energy dependence of a gamma ray should average the energy dependences of these virtual particles. There are many possible processes, i.e. inverse pion decay, the Compton effect, pair production, etc. An average mass called $m_{ray}$ is used in (\ref{Eintro}) and is varied to fit the observed data. The best fit values of $m_{ray}$ and $\phi_{S}$ in Table \ref{table:Fits} are used with (\ref{Eintro}) to obtain the values of $E\gamma_{0}$ and $E\gamma_{\oplus}$ in this table.  \vspace{0.5cm}}
    \label{table:VHEenergy}
    \centering
    \begin{tabular*}{0.5\textwidth}{@{\extracolsep{\fill}}|c|c|c|c|}
      \hline
      \# & $E\gamma_{S}$  (TeV) & $E\gamma_{0}$  (TeV) & $E\gamma_{\oplus}$  (TeV) \\\hline\hline
1. &    0.164 &  0.159 &  0.204   \\\hline
2. &    0.301 &  0.275 &  0.514   \\\hline
3. &   0.509 &  0.417 &  1.25   \\\hline
4. &   0.795 &  0.566 &  2.65    \\\hline
5. &    1.20 &  0.722 &  5.04   \\\hline
6. &    1.80 &  0.895 &  9.12   \\\hline
7. &    2.77 &  1.10 &  16.3   \\\hline
8. &   4.42 &  1.35 &  29.3   \\\hline
9. &    7.35 &  1.66 &  53.6   \\\hline
10. &    12.9 &  2.05 &  100.   \\\hline
11. &   23.3 &  2.54 &  191.   \\\hline
    \end{tabular*}
  \end{table*}

\pagebreak
\clearpage
  \begin{table*}[!h]
    \caption{The CR and VHE gamma ray fluxes at Earth and at the typical SNR for the CRs and gamma rays in Tables \ref{table:CRenergy} and \ref{table:VHEenergy}.  The CR flux at Earth, $dNp/dEp_{\oplus},$ represents experimental data. The CR flux at the SNR follows from the CR flux at Earth and particle state energy dependence, (\ref{dNpdEpSNR}). The gamma ray flux at the SNR is the result of integrating over the CR proton energy at the SNR using the method of Kelner et al. (2006). The gamma ray flux at Earth is found by (\ref{dNgdEgEarth}). The best fit values of $m_{ray}$ and $\phi_{S}$ in Table \ref{table:Fits} are used to obtain the calculated results in columns 3, 4 and 5.  \vspace{0.5cm}}
    \label{table:Flux}
    \centering
    \begin{tabular*}{\textwidth}{@{\extracolsep{\fill}}|c|c|c||c|c|}
      \hline
      \# &  $dNp/dEp_{\oplus}$ & $dNp/dEp_{S}$  & $dN\gamma/dE\gamma_{S}$  & $dN\gamma/dE\gamma_{\oplus}$   \\\hline\hline
1    &       $ 2.18 \times  10^{-13} $  &	 $ 2.43 \times  10^{-13} $   & 	 $ 2.87 \times  10^{-11} $   &	 $ 1.69 \times  10^{-11} $     \\\hline
2    & 	 $ 3.48 \times  10^{-14} $   & 	 $ 4.78 \times  10^{-14} $   & 	 $ 5.23 \times  10^{-12} $   &	 $ 1.88 \times  10^{-12} $     \\\hline
3   & 	 $ 5.56 \times  10^{-15} $   &	 $ 1.08 \times  10^{-14} $   &	 $ 1.15 \times  10^{-12} $   &	 $ 2.73 \times  10^{-13} $     \\\hline
4   &  	 $ 8.87 \times  10^{-16} $   &	 $ 2.49 \times  10^{-15} $   & 	 $ 2.92 \times  10^{-13} $   &	 $ 5.37 \times  10^{-14} $     \\\hline
5   & 	 $ 1.41 \times  10^{-16} $   &	 $ 5.52 \times  10^{-16} $   & 	 $ 7.70 \times  10^{-14} $   &	 $ 1.21 \times  10^{-14} $     \\\hline
6   & 	 $ 2.25 \times  10^{-17} $   & 	 $ 1.14 \times  10^{-16} $   & 	 $ 1.90 \times  10^{-14} $   &	 $ 2.69 \times  10^{-15} $     \\\hline
7   &  	 $ 3.58 \times  10^{-18} $   &	 $ 2.19 \times  10^{-17} $   & 	 $ 4.11 \times  10^{-15} $   &	 $ 5.38 \times  10^{-16} $     \\\hline
8   & 	 $ 5.70 \times  10^{-19} $   & 	 $ 3.98 \times  10^{-18} $   & 	 $ 7.70 \times  10^{-16} $   &	 $ 9.55 \times  10^{-17} $     \\\hline
9   & 	 $ 9.06 \times  10^{-20} $   & 	 $ 6.93 \times  10^{-19} $   & 	 $ 1.33 \times  10^{-16} $   &	 $ 1.58 \times  10^{-17} $     \\\hline
10   & 	 $ 1.44 \times  10^{-20} $   & 	 $ 1.17 \times  10^{-19} $   & 	 $ 2.23 \times  10^{-17} $   &	 $ 2.59 \times  10^{-18} $     \\\hline
11   &   $ 2.27 \times  10^{-21} $   & 	 $ 1.92 \times  10^{-20} $   & 	 $ 3.70 \times  10^{-18} $   &	 $ 4.22 \times  10^{-19} $  \\\hline
    \end{tabular*}
  \end{table*}

\pagebreak

\appendix

\pagebreak
\section{Problems} \label{problems}

[Answers can be found in this section in the LaTeX source document hidden by \%s.]

\vspace{0.3cm}
\noindent 1.  (a) Use the values of $Ep_{0}$ listed in Table \ref{table:CRenergy} to calculate $Ep_{\oplus}$ and $Ep_{S}$ using formula (\ref{Eintro}). Find the percent differences with the values of $Ep_{\oplus}$ and $Ep_{S}$ found using (\ref{E1}). (b) Similarly, for the $E\gamma_{0}$ listed in Table~\ref{table:VHEenergy}, find the percent differences of the values of $E\gamma_{\oplus}$ and $E\gamma_{S}$ found by (\ref{Eintro}) and (\ref{E1}).

\vspace{0.3cm}
\noindent 2. Wherever needed, use the best fit values from Table \ref{table:Fits} in this problem. (a) Use the energies listed in Table \ref{table:CRenergy}, the fluxes in Column 2 of Table \ref{table:Flux} and (\ref{dNpdEpSNR}) to verify the CR fluxes at the SNR, $dNp/dEp_{S}$ in column 3 of Table \ref{table:Flux}. (b) Likewise, use (\ref{dNgdEgEarth}) to verify the $dN\gamma/dE\gamma_{\oplus}$ listed in Column 5 of Table \ref{table:Flux}.

 



\vspace{0.3cm}
\noindent 3. The 11 gamma ray energies $E\gamma_{S}$ in Table \ref{table:VHEenergy} are each 17\% of the 11 CR energies $Ep_{S}$ in Table \ref{table:CRenergy} in keeping with the $\delta$-functional approach that assumes secondary gamma ray energies are a fixed fraction, $\kappa$ = 0.17, of incident CR proton energies. Given a proton flux at the SNR, $dNp/dEp_{S},$ from Table \ref{table:Flux} and the formula for the inelastic cross section $\sigma_{inel},$ Eq. (79) from Kelner et al.\cite{Kelner}, with the normalization chosen to cancel the factors $c$ and $n_{H},$ one can obtain gamma ray fluxes $dN\gamma/dE\gamma_{S}$ using the $\delta$-functional approach. One finds that
$$ \left[\frac{dN_{\gamma}}{dE\gamma_{\oplus}}\right]_{\delta-functional} =  A c n_{H} \kappa \sigma_{inel}(Ep_{S}) \frac{dNp}{dEp_{S}} \, , $$
where the normalization can be adjusted to match the values in Table \ref{table:Flux} (I use $A c n_{H}$ = 20.) Compare the resulting values with the ones listed in Table \ref{table:Flux} which are found by integration, see Eq. (71) from Kelner et al.\cite{Kelner} 



\vspace{0.3cm}
\noindent 4. The observed CR spectrum in the region of interest between the dashed lines in Fig. 1 is approximately a power law, $dNp/dEp_{\oplus}$ = $Ap_{\oplus} Ep_{\oplus}^{2.65},$ where $Ap_{\oplus}$ is a constant. (a) Find the range of the parameter $Ep_{0}$ in the region of interest. (b) Using $Ep_{0}$ as a parameter, or otherwise, find an expression for the CR spectrum at the SNR, $dNp/dEp_{S}.$ (c) Use the $\delta-$functional method, described briefly in Problem 3, to obtain an expression for the VHE gamma ray spectrum at the SNR, $dN\gamma/dE\gamma_{S}.$ [It may be useful to use $E\gamma_{0}$ as a parameter and find the range of $E\gamma_{0}$ that corresponds to the range of $E\gamma_{S}$ (= $\kappa Ep_{S}$).] (c) Find an expression for the VHE gamma ray spectrum at the Earth. How does this result compare with the result using the analysis of Kelner (2006) displayed in Fig. 7?


\vspace{0.3cm}
\noindent 5. Can you give an explanation of why the best fit mass $m_{ray}$ is about half of a neutral pion mass?

%

\end{document}